\documentclass[12pt]{article}
\usepackage{amsmath}
\def\a{\alpha}
\def\b{\beta}

\def\d{\delta}
\def\e{\epsilon}

\def\g{\gamma}
\def\p{\psi}

\def\n{\nu}

\def\t{\theta}
\def\r{\rho}
\def\s{\sigma}

\def\T{\Theta}

\def\vf{\varphi}

\def\be{\begin{equation}}
\def\ee{\end{equation}}
\def\arr{\begin{array}{rll}}
\def\ea{\end{array}}
\def\bea{\begin{eqnarray}}
\def\eea{\end{eqnarray}}

\def\N2{$N{=}2$}

\def\>{\rangle}
\def\<{\langle}
\def\+{\dagger}
\def\={\ =\ }

\begin{document}
\renewcommand{\thefootnote}{\fnsymbol{footnote}}
\begin{titlepage}
\setcounter{page}{0}
\begin{flushright}
LMP-TPU--7/08  \\
\end{flushright}
\vskip 1cm
\begin{center}
{\LARGE\bf Particle dynamics on $AdS_2 \times S^2$ background}\\
\vskip 0.5cm
{\LARGE\bf with two--form flux}\\
\vskip 2cm
$
\textrm{\Large Anton Galajinsky\ }
$
\vskip 0.7cm
{\it
Laboratory of Mathematical Physics, Tomsk Polytechnic University, \\
634050 Tomsk, Lenin Ave. 30, Russian Federation} \\
{Email: galajin@mph.phtd.tpu.edu.ru}

\end{center}
\vskip 1cm
\begin{abstract} \noindent
Different aspects of particle dynamics on $AdS_2 \times S^2$ background with two--form flux are discussed.
These include solution of equations of motion, a canonical transformation
to conformal mechanics and an $N=4$ supersymmetric extension.
\end{abstract}

\vspace{0.5cm}

PACS: 04.60.Ds; 11.30.Pb; 12.60.Jv\\ \indent
Keywords: particle mechanics, Bertotti-Robinson space, $N=4$ supersymmetry
\end{titlepage}

\renewcommand{\thefootnote}{\arabic{footnote}}
\setcounter{footnote}0

\noindent
{\bf 1. Introduction}\\

Over the past few years there has been an upsurge of interest in the model of a
relativistic particle propagating near the horizon of the
extreme Reissner-Nordstr\"om black hole \cite{kallosh}--\cite{ast}.
The near horizon geometry in this case corresponds to the $AdS_2 \times S^2$ space--time
with two--form flux. A peculiar feature
of this system is that it admits two dual descriptions known in the literature as the $AdS$ and
conformal bases.

Originally, it was demonstrated in \cite{kallosh} that in the limit
when the black hole mass $M$ is large, the difference between the particle mass and the absolute value of its charge
$(m-|e|)$ tends to zero with $M^2 (m-|e|)$ kept fixed, one recovers the conventional
$d=1$ conformal mechanics of \cite{aff}.
In particular, the absence of a normalizable ground state in the conformal mechanics
and the necessity to redefine the Hamiltonian \cite{aff} were given a new black hole
interpretation \cite{kallosh}.  Notice that the angular variables effectively decouple in the
aforementioned limit and show
up only in an indirect way via the effective coupling constant characterizing the conformal mechanics.

Later on it was argued in \cite{ikn,bik} that, discarding the angular variables, a particle
on $AdS_2$ background and the conformal mechanics of \cite{aff} can be related by an invertible coordinate transformation.
In contrast to \cite{kallosh}, the connection holds for any finite value of the black hole mass and
has its origin in the possibility to choose different cosets of the conformal group $SO(2,1)$ within the method
of nonlinear realizations  \cite{ikn}. A proper extension of the $d=1$ conformal mechanics by angular degrees of freedom
which is equivalent to a massive charged particle on $AdS_2 \times S^2$ background with
two--form flux was constructed in \cite{bgik} (see also a related work \cite{lp}).
In particular, a simple canonical transformation was found which directly relates symmetry
generators (including the Hamiltonian) in both the pictures. As the transformation is invertible,
different aspects of dynamics in one model can be studied in terms of the other and vice versa.
The two pictures were called the $AdS$ and conformal bases.

The extreme Reissner-Nordstr\"om black hole solution of Einstein--Maxwell theory can be embedded into $d=4$, $N=2$
supergravity by adding two gravitini (for a review see e.g. \cite{moh}). As in the near horizon limit
there is an enhancement of symmetry, for the particle on $AdS_2 \times S^2$ background one can construct an $N=4$
supersymmetric extension. The action of the corresponding super $0$--brane was found in \cite{zhou1,zhou}
with the use of the supercoset approach. Notice, however, that a consistent gauge fixed Hamiltonian formulation
in terms of physical variables obeying canonical commutation relations is unknown.

Most of the developments mentioned above were focused on the case when a magnetic charge of the
extreme Reissner-Nordstr\"om black hole vanishes. As the presence of a magnetic charge causes
essential change in particle dynamics (see e.g. \cite{ply}), it is interesting to see
which is the conformal model in this case and how the
correspondence between the $AdS$ and conformal descriptions is altered.

The purpose of this work is to extend the analysis in \cite{bgik} to the case of a nonvanishing magnetic charge.
In the next section we briefly discuss the geometry of background fields. In sect. 3 particle dynamics on
$AdS_2 \times S^2$ background with two--form flux is analyzed within the Hamiltonian formalism. The conserved charges
are found which allow us to integrate the equations of motion in an efficient way. A conformal picture is considered
in sect. 4. An extension of the conformal mechanics \cite{aff} by angular variables is given which is
characterized by two independent coupling constants. Making use of the rotation invariance, we
construct a simple canonical transformation which relates the $AdS$ and conformal bases. An $N=4$ supersymmetric extension
of the system in the conformal picture is discussed in sect. 5. Making use of the Hamiltonian methods
we arrive at an on--shell component formulation for the $D(2,1;\a)$--invariant mechanics of \cite{ikl} with $\a=-1$.
It is interesting to note that in order to accommodate $N=4$ supersymmetry in the original bosonic conformal
mechanics one has to identify the two couplings. Sect. 6 is devoted to an $N=4$ supersymmetric
generalization of the model in the $AdS$ basis. In particular, we construct a new Hamiltonian formulation in terms
of physical variables which obey canonical commutation relations. We summarize the results in sect. 7.
Our conventions for dealing with $SU(2)$--spinors and the commutation relations of $d=1$, $N=4$ superconformal algebra
are given in Appendix.

\vspace{0.5cm}

\noindent
{\bf 2. Geometry of background fields}\\

Our starting point is the extreme Reissner-Nordstr\"om black hole solution of Einstein--Maxwell equations
(for a review see e.g. \cite{moh})
\be
{d s}^2=-{\left(1-\frac{M}{r}\right)}^2 dt^2+{\left(1-\frac{M}{r}\right)}^{-2} dr^2+r^2 d\Omega^2\ , \quad
A=-\frac{q}{r} dt+p \cos\t d\vf\ .
\ee
Here $M$, $q$, $p$ are the mass, the electric and magnetic charges, respectively, and
$d\Omega^2=d\t^2+\sin^2\t d\vf^2$ is the standard metric on a sphere. For the extreme solution one has $M=\sqrt{q^2+p^2}$.
Throughout the paper we use units for which $G=1$.

The near horizon limit is most easily accessible in isotropic coordinates ($r~\rightarrow ~r-M$)
which cover the region outside the horizon only
\be\label{RN1}
{d s}^2=-{\left(1+\frac{M}{r}\right)}^{-2} dt^2+{\left(1+\frac{M}{r}\right)}^2 \left( dr^2+r^2 d\Omega^2\right)\ .
\ee
When $r\rightarrow 0$ the metric takes the form
\be\label{BR}
{d s}^2=-{\left(\frac{r}{M}\right)}^2 dt^2+{\left(\frac{M}{r}\right)}^2 dr^2+M^2 d\Omega^2\ ,
\ee
while implementing the limit in the two--form field strength, one finds the background vector field
\be\label{Pot}
\quad
A=\frac{q}{M^2} r dt+p \cos\t d\vf\ .
\ee
The last two lines give the Bertotti-Robinson solution of Einstein--Maxwell equations.

Notice that
in the literature on the subject it is customary to use other coordinates where the horizon is at $r=\infty$.
In particular, the use of these coordinates facilitates the analysis in \cite{kallosh}. In this paper we refrain from
using such a coordinate system.

From (\ref{BR}) it follows that in the near horizon limit the space--time geometry is the product
of a two-dimensional sphere of radius $M$ and a two-dimensional pseudo Riemannian space--time with the metric
\be\label{ads2}
{d s}^2=-{\left(\frac{r}{M}\right)}^2 dt^2+{\left(\frac{M}{r}\right)}^2 dr^2\ .
\ee
The latter proves to be the metric of $AdS_2$. In order to see this, consider the
hyperboloid in $\mathcal{R}^{2,1}$
\be
-\eta_{AB}x^A x^B=M^2\ , \qquad \eta_{AB}=\mbox{diag} (-,+,-)\ ,
\ee
parameterized by the Poincar\'e coordinates $(t,r)$
\be\label{hyp}
x^0=\frac{1}{2r} (1+r^2(M^2-t^2)), \quad x^1=\frac{1}{2r} (1-r^2(M^2+t^2)), \quad x^2=Mrt\ .
\ee
Since $x^0-x^1>0$, the local coordinates cover only half of the hyperboloid\footnote{In order to avoid closed time--like curves,
one considers the universal covering of the hyperboloid with $-\infty<t<\infty$, $0<r$.}. Calculating the metric $ds^2=\eta_{AB} dx^A dx^B$ induced on the surface
(\ref{hyp}) and making the shift $r \rightarrow M^2 r$, one gets precisely (\ref{ads2}). Notice that in this picture the black hole mass $M$ is equal to the radius of $S^2$ ($AdS_2$). It is worth mentioning also that, by construction, the isometry group of the metric (\ref{BR}) is
$SO(2,1)\times SO(3)$.

To summarize, the background geometry is that of the $AdS_2 \times S^2$ space--time with 2--form flux.
\vspace{0.5cm}

\noindent
{\bf 3. Particle dynamics on $AdS_2\times S^2$}\\

Having fixed the background fields, we then consider the action of a relativistic particle
on such a background
\be\label{start}
S=-\int d t \left(m\sqrt{ {(r/M)}^2-{(M/r)}^2 {\dot r}^2 -M^2 ({\dot\t}^2
+\sin^2 \t {\dot\vf}^2)   }+eqr/M^2 +e p \cos \t \dot\vf~\right)\ .
\ee
Here $m$ and $e$ are the mass and the electric charge of a particle, respectively.

The particle dynamics is most easily analyzed within the Hamiltonian formalism. Introducing the momenta $(p_r,p_\t,p_\vf)$ canonically
conjugate to the configuration space variables $(r,\t,\vf)$, one finds the Hamiltonian
\be\label{h}
H=(r/M)\left(\sqrt{m^2+{(r/M)}^2 p_r^2 +{(1/M)}^2(p_\t^2+\sin^{-2}\t {(p_\vf+e p \cos\t)}^2)} +e q/M\right)\ ,
\ee
which generates  time translations.
In agreement with the isometries of the background metric one also finds the conserved quantities
\bea\label{kd}
&&
K=M^3/r \left(\sqrt{m^2+{(r/M)}^2 p_r^2 +{(1/M)}^2(p_\t^2+\sin^{-2}\t {(p_\vf+e p \cos\t)}^2)} -e q/M\right)+
\nonumber\\[2pt]
&& \qquad
+t^2 H+2tr p_r\ , \qquad D=tH+r p_r\ ,
\eea
which generate special conformal transformations and dilatations, respectively. Together with the Hamiltonian they form
$so(2,1)$ algebra
\be\label{confalg}
\{H,D \}=H\ , \quad \{H,K \}=2D\ , \quad \{D,K \} =K\ .
\ee

The generators of rotations
\bea\label{su2}
&&
J_1=-p_\vf \cot \t \cos \vf  -p_\t \sin \vf-e p \cos \vf \sin^{-1}\t\ ,
\nonumber\\[2pt]
&&
J_2=-p_\vf \cot \t \sin \vf +p_\t \cos \vf -e p \sin \vf \sin^{-1}\t\ ,
\nonumber\\[2pt]
&&
J_3=p_\vf\ , \qquad   \{ J_a,J_b \}=\epsilon_{abc} J_c\ , \qquad \quad\epsilon_{123}=1\ .
\eea
enter the Hamiltonian via the Casimir element
\be\label{casimir}
J^a J_a=p_\t^2+\sin^{-2}\t {(p_\vf+e p \cos\t)}^2+{(e p)}^2\ ,
\ee
and, hence, are conserved due to $su(2)$ algebra they form.

When analyzing solutions of equations of motion, two distinct cases should be examined.
First consider the situation when the magnetic charge of the black hole vanishes
\be
p=0\ , \quad M=|q|\ .
\ee
In this case the particle moves on a plane
orthogonal to the angular momentum vector $J_i$. Making use of the rotation invariance one can choose the reference frame where $J_i$
is along $x^3$-axis, i.e.
\be
\t=\pi/2\ , \quad  p_\t=0\ \quad \rightarrow \quad J_1=0, \quad J_2=0, \quad J_3=p_\vf=L\ ,
\ee
with $L$ a constant\footnote{We assume that $L \ne 0$. When $L=0$ the particle travels towards the horizon at $r=0$ along a straight line.}.
Then from the conservation laws (\ref{h}) and (\ref{kd}) one can fix the dynamics of the radial coordinate
\be\label{r}
r(t)=\frac{E M^2}{\sqrt{a^2 (t)+b^2}+c}\ , \qquad p_r (t)=\frac{a(t)(\sqrt{a^2 (t)+b^2}+c)}{E M^2}\ ,
\ee
where $E=H$ is the energy and we abbreviated
\be\label{coef}
a(t)=D-t E\ , \qquad b^2=m^2 M^2+L^2\ , \qquad c=e q\ .
\ee
The evolution of the angular variable is found by a straightforward integration
\be\label{p}
\vf(t)=-\frac{L}{\sqrt{b^2-c^2}} \left( \arctan \frac{a(t)}{\sqrt{b^2-c^2}}-\arctan \frac{c a(t)}{\sqrt{a^2 (t)+b^2}\sqrt{b^2-c^2}} \right)+\vf_0\ ,
\quad p_\vf(t)=L\ .
\ee

It is important to notice that the conserved charges (\ref{h}), (\ref{kd}) also specify the value of the Casimir element of $so(2,1)$ algebra
realized in the model in terms of the parameters of the particle and those of the background
\be
E K -D^2=b^2-c^2=M^2 (m^2-e^2) +L^2\ .
\ee
This should correlate with the bound
$b^2-c^2>0$
revealed by the explicit solution given above. The latter also assures that the energy of the particle $E=(r/{M}^2) (\sqrt{a^2 (t)+b^2}+c)$
is positive even if $c$ is negative. Indeed, if $c<0$ then from the condition $b^2-c^2>0$
one immediately gets
\be\label{au}
(\sqrt{a^2 (t)+b^2}+c)(\sqrt{a^2 (t)+b^2}-c)>0\ ,
\ee
which means that the first factor entering (\ref{au}) is positive.

As $\dot r$ is proportional to $a(t)$ with a positive coefficient, depending on the initial data, the particle either
goes directly towards the black hole horizon located at $r=0$, or it moves away for some time, slows down
with the turning point at $t=D/E$, and then travels back towards $r=0$.
The orbit looks particularly simple when the particle is electrically neutral, i.e. $c=0$
\be
r(\vf)=\frac{E M^2}{b} |\cos \left( b(\vf-\vf_0)/L\right)| \ .
\ee
The trajectory is a kind of a loop which starts and ends at $r=0$ and has a symmetry axis typical for
rotation invariant systems.

Now consider the case when the magnetic charge $p$ of the black hole does not vanish. In this case the
particle moves on the cone (turning to Cartesian coordinates)
\be\label{cone}
\frac{x^i J_i}{\sqrt{x^2}}=-ep\ .
\ee
As before, one can use the rotation invariance so as to pass to the reference frame where $J_i$
is along $x^3$-axis. This specifies the canonical pair $(\t,p_\t)$
\be
J_1=0, \quad J_2=0, \quad J_3=p_\vf=L\  \quad \rightarrow \quad \cos\t=-ep/L\ , \quad  p_\t=0\ ,
\ee
and imposes the natural bound $|\frac{ep}{L}|\le 1$. The solutions of equations of motion for $(r(t),p_r (t))$ and $(\vf(t), p_\vf (t))$ prove
to maintain their previous form
(\ref{r}), (\ref{p}) with $a(t)$ and $c$ unchanged, but $b^2$ modified
\be
b^2=m^2 M^2+L^2-{(ep)}^2\ .
\ee
For $p \ne 0$ the qualitative behavior of a particle is similar to the previous case but this time it is confined to
move on the cone (\ref{cone}).

\vspace{0.5cm}

\noindent
{\bf 4. A relation to conformal mechanics}\\

The conventional conformal mechanics in one dimension is governed by the action functional \cite{aff}
\be\label{act}
S=\frac 12 \int d t \left({\dot x}^2 - \frac{g}{x^2} \right)\ ,
\ee
where $g$ is the coupling constant. Passing to the Hamiltonian formalism one finds the conserved charges
\be\label{ham1}
H'=\frac{p^2}{2} + \frac{g}{2 x^2}\ , \quad
D'=tH'-\frac 12 x p, \quad K'=t^2 H'-t(x p) +\frac 12 x^2 \ ,
\ee
which altogether form $so(2,1)$ algebra (\ref{confalg}). Guided by this observation, the authors of \cite{kallosh}
argued that the quantum mechanics of a test particle moving near the horizon of the extreme Reissner-Nordstr\"om black
hole\footnote{In \cite{kallosh} only the case of the vanishing magnetic charge was discussed.}
matches the old conformal mechanics (\ref{act}) in the limit
\be
M\rightarrow \infty\ , \qquad  (m-|e|)\rightarrow 0\ ,
\ee
with $M^2(m-|e|)$ fixed.

In \cite{bgik} the conformal mechanics (\ref{act}) was extended by a couple of angular variables in such a way that
the resulting model is related to a particle moving near the horizon of the extreme Reissner-Nordstr\"om black hole
by a canonical transformation (for a related work see \cite{lp}). The construction in \cite{bgik} does not appeal to any specific limit and is valid for any finite value of the black hole mass.

In this section we generalize the analysis in \cite{bgik} to the case when a test particle couples to the magnetic charge of
the black hole. As compared to the calculation in \cite{bgik}, the use of the rotation invariance notably facilitates the
analysis.

Consider a specific extension of the model (\ref{act}) by two angular degrees of freedom $(\Theta,\Phi)$
\be\label{act1}
S=\frac 12 \int d t \left( {\dot x}^2 +\frac 14 x^2 ({\dot\Theta}^2
+\sin^2 \Theta {\dot\Phi}^2)- \frac{g}{ x^2}-2 \nu \cos\Theta \dot\Phi \right)\ ,
\ee
where $\nu$ is a new coupling constants and $x$ is now treated as a radial coordinate in the enlarged configuration space.
This theory arises, in particular, in the bosonic limit
of the superconformal mechanics associated with the supergroup $D(2,1;\a)$ for $\a=-1$ \cite{ikl}.
Notice that in \cite{ikl} the $g$ and $\nu^2$ couplings were identified (see also the discussion in sect. 5).
In non--supersymmetric case they are independent.

That the new degrees of freedom do not destroy the conformal symmetry of the original model is most easily verified within the Hamiltonian
formalism. Indeed, given the Hamiltonian
\be\label{ham2}
H'=\frac{p^2}{2} + \frac{g}{2x^2}+\frac{2}{x^2} (p_\T^2+\sin^{-2} \T {(p_\Phi+\nu\cos\T)}^2)\ ,
\ee
where $(p,p_\T,p_\Phi)$ designate momenta canonically conjugate to $(x,\T,\Phi)$,
the generators of dilatations $D'$ and special conformal transformations $K'$ are constructed following the prescription
(\ref{ham1}) and the full algebra proves to be $so(2,1)$.

As might be anticipated from the form of the action (\ref{act1}), the system accommodates rotation invariance.
The corresponding generators are derived from (\ref{su2}) by the obvious change of the canonical pairs
$(\vf,p_\vf)\rightarrow (\Phi,p_\Phi)$,
$(\t,p_\t)\rightarrow (\Theta,p_\T)$
and the coupling constants
$ep\rightarrow \nu$. They are trivially conserved because the
angular variables enter the Hamiltonian via the Casimir element of $so(3)$ algebra realized in the model.

Now let us demonstrate that the system (\ref{ham2}) and a particle on $AdS_2 \times S^2$ background with $2$--form flux are
related by a canonical transformation. In order to simplify the analysis, let us use the rotation invariance intrinsic to both
the models and pass for each system to the reference frame where the conserved angular momentum vector is along
$x^3$--axis\footnote{Our construction implies  $p_\Phi=p_\vf=L$ on--shell.}.
This allows one to disregard
the pairs $(\t,p_\t)$, and $(\T,p_\T)$
\bea
&&
\cos\t=-ep/L\ , \quad  p_\t=0\ ,
\nonumber\\[2pt]
&&
\cos\T=-\nu/L\ , \quad  p_\T=0\ .
\eea
Following \cite{bgik}, we then search for a canonical transformation which brings the symmetry generators characterizing the model
(\ref{start}) precisely to those of the system (\ref{act1}). Comparing the conserved charges (including the Hamiltonian) in both the pictures,
one immediately finds the desired transformation
\bea\label{trans}
&&
x={ \left[\frac{2M^2}{r} \left(\sqrt{m^2 M^2+{(r p_r)}^2 +{(L p_\vf -{(ep)}^2)}^2 /(L^2-{(ep)}^2) } -eq \right)\right]}^{\frac 12}\ ,
\nonumber\\[2pt]
&&
p=-2r p_r { \left[\frac{2M^2}{r} \left(\sqrt{m^2 M^2+{(r p_r)}^2 +{(L p_\vf -{(ep)}^2)}^2 /(L^2-{(ep)}^2) } -eq \right)\right]}^{-\frac 12}\ ,
\nonumber\\[2pt]
&&
p_\Phi=p_\vf \ .
\eea
The corresponding Poisson brackets prove to be canonical. When establishing this correspondence, one has to specify
the couplings of the conformal mechanics in terms of the parameters characterizing the particle on $AdS_2\times S^2$
\be\label{const}
\nu=ep\ ,\qquad g=4(m^2 M^2-{(eq)}^2) \ .
\ee

A transformation law of the last missing variable
$\Phi$ is then found with the help of
(\ref{trans}). Imposing the canonical relations
\be\label{rels}
\{\Phi,x \}=0\ , \qquad \{\Phi,p\}=0\ , \qquad  \{\Phi,p_\Phi\}=1\ ,
\ee
which are to be calculated with respect to the variables $(r,p_r)$, $(\vf,p_\vf)$, and taking the ansatz
\be\label{transf}
\Phi=\vf+A(s,p_\vf)\ ,
\ee
with $s=(rp_r)$ and $A(s,p_\vf)$ an arbitrary function, one reduces (\ref{rels}) to a single
ordinary differential equation. This yields the solution
\be
A(s,p_\vf)=-\frac{\a}{\sqrt{k^2-c^2}} \left(\arctan \frac{s}{\sqrt{k^2-c^2}}+\arctan \frac{cs}{\sqrt{k^2-c^2}\sqrt{k^2+s^2}} \right)\ ,
\ee
where we denoted
\be
\a=\frac{L(L p_\vf-{(ep)}^2)}{(L^2-{(ep)}^2)}\ , \quad k^2=m^2 M^2+\frac{{(L p_\vf -{(ep)}^2)}^2}{(L^2-{(ep)}^2)}\ , \quad c=eq\ .
\ee

In the consideration above we made explicit use of the rotation invariance. Obviously, rotation is a canonical transformation. So,
the transformation relating the models (\ref{h}) and (\ref{ham2}) is a superposition of (\ref{trans}), (\ref{transf})
and two rotations. The latter affect angular variables only. Then it is not hard to guess the radial part of the transformation
\bea\label{new}
&&
x={ \left[\frac{2M^2}{r} \left(\sqrt{m^2 M^2+{(r p_r)}^2 +p_\t^2+\sin^{-2}\t {(p_\vf+e p \cos\t)}^2 } -eq \right)\right]}^{\frac 12}\ ,
\nonumber\\[2pt]
&&
p=-2r p_r { \left[\frac{2M^2}{r} \left(\sqrt{m^2 M^2+{(r p_r)}^2 +p_\t^2+\sin^{-2}\t {(p_\vf+e p \cos\t)}^2 } -eq \right)\right]}^{-\frac 12},
\eea
which indeed brings (\ref{ham2}) to (\ref{h}), provided the identification of the couplings (\ref{const}) holds.
Notice that the momentum squared is invariant
\be\label{new1}
p_\T^2+\sin^{-2} \T {(p_\Phi+\nu\cos\T)}^2=
p_\t^2+\sin^{-2}\t {(p_\vf+e p \cos\t)}^2\ .
\ee
Although in this picture the transformation of the angular variables appears to be rather complicated\footnote{
For the case of the vanishing magnetic charge it was found in \cite{bgik} by identifying the $so(3)$ generators in both the pictures.},
for practical uses one does not need to know
its explicit form and the relations (\ref{new}) and  (\ref{new1}) prove to be sufficient.

To summarize, the canonical transformation exposed above establishes the equivalence relation
between the charged massive particle moving near the horizon of the extreme Reissner-Nordstr\"om black hole, which
has a non--vanishing magnetic charge, and the conformal mechanics~(\ref{act1}). Different aspects of dynamics in one model can
be studied in terms of the other and vice versa. It is noteworthy that the equivalence
holds for any fixed value of the black hole mass and does not refer to any specific limit. This is to be contrasted with
the consideration in \cite{kallosh}.

\vspace{0.5cm}

\noindent
{\bf 5. $N=4$ supersymmetric extension in the conformal basis}

\vspace{0.5cm}

An $N=4$ supersymmetric generalization of the model (\ref{act1}) was constructed in \cite{ikl} with the use of the
method of nonlinear realizations\footnote{For an earlier work on $N=2$ model see \cite{Akulov}.}.
The corresponding action was given in terms of superfields. At the component level it involves
non--dynamical auxiliary fields needed for the off-shell closure of the $d=1$, $N=4$ superconformal algebra realized in the model.
Notice that the $N=4$ supersymmetry makes one to identify the $g$ and $\nu^2$ couplings in (\ref{act1}).

Aiming at the construction of an $N=4$ supersymmetric extension of the particle moving near the horizon of the extreme
Reissner-Nordstr\"om black hole, in this section we discuss an $N=4$ supersymmetric generalization of the bosonic conformal mechanics
(\ref{act1}) using the alternative Hamiltonian formalism. The advantage of this approach is that it automatically yields
an on--shell component formulation free from non--dynamical auxiliary fields and offers technical simplifications as
compared to the method in \cite{ikl}. Besides, it can be readily applied to construct important multi--particle superconformal systems,
including the $N=4$ superconformal Calogero model (see e.g. \cite{bgl}--\cite{l}), while the superfield 
appears to be more involved \cite{bks}.
For the particular case of the vanishing coupling constants $g=\nu=0$ the construction
was realized in \cite{bgik}.

Within the framework of the Hamiltonian formalism the construction of an $N=4$ supersymmetric generalization of the system  (\ref{ham2})
amounts to extending the bosonic phase space by a pair of canonically conjugate $SU(2)$--spinors $\p_\a, \bar\p^\a$ (for our
conventions see Appendix) and building in the enlarged phase space a representation of $su(1,1|2)$ superalgebra. Along with the
$so(2,1)$--, and $su(2)$--generators which comprise bosonic symmetries of the model (\ref{ham2}), the superalgebra involves the supersymmetry
generators $Q_\a, \bar Q^\a$ and the superconformal ones $S_\a, \bar S^\a$ (the commutation relations are given in
Appendix). The conditions that the Poisson bracket of $Q_\a$ and $\bar Q^\b$ yields the Hamiltonian and that $Q_\a$ anticommutes with
itself prove to be strong enough to fix the form of the supercharges
\be
Q_\a=p \p_\a+\frac{2i}{x} {(\s^a \p)}_\a J_a +\frac{i}{2x} \bar\p_\a \p^2\ , \qquad
\bar Q^\a =p \bar\p^\a-\frac{2i}{x} {(\bar\p \s^a)}^\a J_a +\frac{i}{2x} \p^\a \bar\p^2\ ,
\ee
and the extended Hamiltonian
\be\label{ham3}
H=\frac{p^2}{2}+\frac{2}{x^2} J^a J_a-\frac{2}{x^2} (\bar\p \s^a \p) J_a +\frac{1}{4x^2} \p^2 \bar\p^2\ .
\ee
Here $J_a$ are the bosonic $su(2)$--generators which are realized as in
(\ref{su2}) with the obvious change $(\vf,p_\vf)\rightarrow (\Phi,p_\Phi)$,
$(\t,p_\t)\rightarrow (\Theta,p_\T)$, $ep\rightarrow \nu$. Comparing this Hamiltonian with (\ref{ham2}) and taking into account
(\ref{casimir}), one concludes that the supersymmetric extension is only possible if one identifies
the $g$ and $\nu$ couplings as follows
\be\label{identi}
g={(2\nu)}^2\ .
\ee
This is in full agreement with the superfield considerations in \cite{ikl}.

As is obvious from (\ref{ham3}), the extended system maintains the conformal symmetry.
Given the Hamiltonian, the generators of dilatations and
special conformal transformations are constructed in the standard way
\be
D=tH-\frac 12 x p, \quad K=t^2 H-t(x p) +\frac 12 x^2 \ .
\ee
Then the Poisson brackets of $K$ with $Q_\a$ and $\bar Q^\a$ yield the superconformal generators
\be
S_\a=x \p_\a -t Q_\a \ , \qquad \bar S^\a=x \bar\p^\a -t \bar Q^\a\ .
\ee

It remains to be discussed the $su(2)$ symmetry realized in the extended model (\ref{ham3}). As the fermionic degrees
of freedom transform as $SU(2)$--doublets, the bosonic generator $J_a$ must be extended so as to include a
piece responsible for the fermions. One can either guess its form or just
calculate the Poisson bracket of $Q_\a$ with $\bar S^\b$
\be
J_a \quad \rightarrow \quad \mathcal{J}_a=J_a+\frac 12 (\bar\p \s_a \p)\ .
\ee

Finally, it is straightforward to check that the generators introduced above do form a
representation of $su(1,1|2)$ superalgebra. The explicit verification makes heavy use of
the properties of the Pauli matrices and spinor rearrangement rules given in Appendix.

In order to construct a Lagrangian formulation reproducing the Hamiltonian (\ref{ham3}), we first notice that
within the Hamiltonian formalism the canonical bracket $\{ \p_\a, \bar\p^\b \}=-i{\d_\a}^\b$ is conventionally understood as
the Dirac bracket
\be
{\{A,B \}}_D=\{A,B \}-i\{A,\chi^\a \}\{\bar\chi_\a,B \}-i\{A,\bar\chi_\a \}\{\chi^\a,B \}\
\ee
associated with the fermionic second class constraints
\be\label{cons}
\chi^\a={p_\p}^\a-\frac i2 \bar\p^\a=0\ , \qquad \bar\chi_\a=p_{\bar\p \a}-\frac i2 \p_\a=0\ .
\ee
Here $({p_\p}^\a,p_{\bar\p \a})$ stand for the momenta canonically conjugate to the variables
$(\p_\a,\bar\p^\a)$, respectively.

Choosing the right derivative for fermionic degrees of freedom, an action functional leading to the
Hamiltonian formulation (\ref{ham3}) is straightforward to build
\bea\label{accon}
&&
S=\int dt \left(\frac 12 {\dot x}^2 +\frac i2 \bar\p^\a {\dot\p}_\a-\frac i2 {\dot{\bar\p}} {}^\a \p_\a
+\frac 18 x^2 ({\dot\Theta}^2
+\sin^2 \Theta {\dot\Phi}^2)- \frac{2 \n^2}{ x^2}-\nu \cos\Theta \dot\Phi
\right.
\nonumber\\[2pt]
&&
\qquad \qquad +
\left.
(\bar\p \s_a \p) \mathcal{L}^a-\frac {3}{4x^2} \p^2 \bar\p^2
 \right)\ .
\eea
Here $\mathcal{L}^a$ is the bosonic part of the angular momentum vector written in configuration space
\bea
&&
\mathcal{L}^1=-\frac 12 \dot\Phi \cos\T \sin \T \cos\Phi -\frac 12 \dot\T \sin\Phi -\frac{2\n}{x^2} \cos\Phi \sin\T\ ,
\nonumber\\[2pt]
&&
\mathcal{L}^2=-\frac 12 \dot\Phi \cos\T \sin\T \sin\Phi+\frac 12 \dot\T \cos\Phi -\frac{2\n}{x^2} \sin\Phi \sin\T\ ,
\nonumber\\[2pt]
&&
\mathcal{L}^3=\frac 12 \dot\Phi \sin^2 \T-\frac{2\n}{x^2}\cos\T\ .
\eea
When relating the action (\ref{accon}) from the Hamiltonian (\ref{ham3}) the spinor identity
\be
(\bar\p \s_a \p)(\bar\p \s_b\p)=-\frac 12 \d_{ab} \p^2 \bar\p^2 \
\ee
proves to be helpful.

Thus, we have constructed an $N=4$ supersymmetric extension of the conformal mechanics
(\ref{act1}) by applying the Hamiltonian methods. The supersymmetry requires the identification (\ref{identi})
of the coupling constants. The model built in this section can be viewed as an on--shell component formulation
for the $D(2,1;\a)$--invariant mechanics of \cite{ikl} with $\a=-1$.
\vspace{0.5cm}

\noindent
{\bf 6. $N=4$ supersymmetric extension in the AdS basis}

\vspace{0.5cm}

Having constructed an $N=4$ supersymmetric extension in the conformal basis, let us
discuss its $AdS$ partner. In the preceding section, when evaluating Poisson brackets of
the $su(1,1|2)$--generators, we used only the canonical relations
\be\label{rela}
\{x,p\}=1\ , \qquad \{J_a,J_b \}=\e_{abc} J_c\ , \qquad \{ \p_\a, \bar\p^\b \}=-i{\d_\a}^\b \ ,
\ee
and the fact that all other brackets involving $(x,p,J_a,\p_a,\bar\p^\a)$ vanish.
Consider the transformation (\ref{new}) relating
the conformal and $AdS$ bases. It
respects (\ref{rela}) provided the $su(2)$--generators in the $AdS$ picture
are taken as in (\ref{su2}).
This is the instance when one does not need to know the explicit form of the canonical
transformation of the angular variables but only their specific combinations, e.g. the $su(2)$--charges.
The fermionic degrees of freedom are kept inert under the transformation from one picture to another.

In order to accommodate $N=4$ supersymmetry in the model regarded  in the conformal picture, one
has to relate the $g$ and $\nu$ couplings as in (\ref{identi}). On the other hand,
the transformation (\ref{new}) to the $AdS$ picture
implies the identification (\ref{const}). Taking into account that for the extreme Reissner-Nordstr\"om black hole
$M^2=q^2+p^2$, one concludes that an $N=4$ supersymmetric extension of the model (\ref{h}) is characterized by the
additional physical requirement
\be
m=|e|\ .
\ee
Thus, the system can be viewed as a BPS superparticle in a BPS background.

Summarizing the above discussion, we can write down the Hamiltonian
\bea\label{h5}
&&
H=\frac{r}{M^2}\left(\sqrt{m^2 M^2+{(r p_r)}^2 +p_\t^2+\sin^{-2}\t {(p_\vf+e p \cos\t)}^2} +e q
\right)
\nonumber\\[2pt]
&&
-\frac{r}{M^2}
((\bar\p \s^a \p) J_a-\frac 18 \p^2 \bar\p^2) {\left(\sqrt{m^2 M^2+{(r p_r)}^2 +p_\t^2+\sin^{-2}\t {{(p_\vf+e p \cos\t)}^2}}-eq\right)}^{-1} \,
\nonumber\\[2pt]
\eea
and the conserved charges of an $N=4$ superparticle propagating on $AdS_2 \times S^2$ background with two--form flux
\bea
&&
K=t^2 H+2tr p_r+\frac{M^2}{r} \left(\sqrt{m^2 M^2+{(r p_r)}^2 +p_\t^2+\sin^{-2}\t {{(p_\vf+e p \cos\t)}^2}}-eq\right)\ ,
\nonumber\\[2pt]
&&
D=tH+r p_r\ ,
\eea
\bea
&&
Q_\a=-\frac{2\left((r p_r) \p_\a-i{(\s^a \p)}_\a J_a -\frac i4 \bar\p_\a \p^2 \right) }
{{ \left(\frac{2M^2}{r} \left(\sqrt{m^2 M^2+{(r p_r)}^2 +p_\t^2+\sin^{-2}\t {(p_\vf+e p \cos\t)}^2 } -eq \right)\right)}^{\frac 12}}\ ,
\nonumber\\[2pt]
&&
S_\a=\p_\a { \left(\frac{2M^2}{r} \left(\sqrt{m^2 M^2+{(r p_r)}^2 +p_\t^2+\sin^{-2}\t {(p_\vf+e p \cos\t)}^2 } -eq \right)\right)}^{\frac 12}
-t Q_\a\ ,
\nonumber\\[2pt]
&&
\mathcal{J}_a=J_a+\frac 12 (\bar\p \s_a \p)\ ,
\eea
where $J_a$ are defined in (\ref{su2}).

A Lagrangian formulation corresponding to the Hamiltonian (\ref{h5}) is straightforward to construct. It proves sufficient to treat
the fermionic degrees of freedom like we did in the preceding section and apply the Legendre transform to the Hamiltonian. However,
a resulting formulation does not literally coincide with the gauge fixed version of the super $0$--brane on $AdS_2 \times S^2$ built within
the Green--Schwarz approach \cite{zhou1,zhou}. A specific field redefinition is to be implemented in order to relate the two systems.
The reason is that the Hamiltonian formulation of the super $0$--brane involves fermionic second class constraints which
depend on the background fields. Introducing the Dirac bracket one can solve the second class constraints and
eliminate the fermionic momenta.
However, the brackets for the remaining physical variables are not canonical. In particular, bosonic variables have nonvanishing brackets
with physical fermions. In general, one has to implement
a nontrivial field redefinition so as to bring the brackets to a canonical form. Notice that
the canonical brackets are also needed for constructing a conventional quantum mechanical description.
The advantage of the model (\ref{h5}) is that the physical variables do obey the canonical relations. So, the system is likely
to describe the Hamiltonian formulation of the gauge fixed super $0$--brane on $AdS_2 \times S^2$ written in proper coordinates.
Finding an explicit form of the field redefinition is an interesting problem which deserves further investigation.

\vspace{0.5cm}

\noindent
{\bf 7. Conclusion}

\vspace{0.5cm}

To summarize, in the present paper we
studied the dynamics of a massive charged particle moving on $AdS_2 \times S^2$ background with
$2$--form flux and constructed its conformal partner. The connection between the two models
is provided by a specific canonical transformation which relates symmetry generators in both the pictures.
An $N=4$ supersymmetric extension of the model in the conformal bases was constructed and then combined with the
canonical transformation so as to produce a new Hamiltonian formulation of an $N=4$ superparticle on
$AdS_2 \times S^2$.

Turning to possible further developments, the first issue is how to generalize the present analysis to the case of
$D(2,1;\a)$ supergroup with $\a\ne -1$. It is interesting to see which background geometry corresponds
to the superconformal particle of \cite{ikl} written in the $AdS$ basis and what is the geometrical
meaning of the parameter $\alpha$. Then it remains to explore how the equivalence established within the Hamiltonian
formalism is translated into the Lagrangian language and how it is linked to the off-shell map of \cite{ikn,bik}.
A more technical issue is to find a field redefinition that relates our Hamiltonian formulation of the $N=4$ superparticle
on $AdS_2 \times S^2$ to the Hamiltonian formulation of the gauge fixed super $0$--brane of \cite{zhou}.

\vspace{0.5cm}

\noindent{\bf Acknowledgements}\\

\noindent
We thank D. Astefanesei and S. Krivonos for useful comments.
This work was supported by RF Presidential grants MD-2590.2008.2,
NS-2553.2008.2, RFBR grant 08-02-90490-Ukr and the DAAD grant A/08/08578.

\vspace{0.5cm}

\noindent
{\bf Appendix}

\vspace{0.5cm}

Throughout the text we use a lower Greek index to designate an $SU(2)$--doublet representation. Complex conjugation
yields an equivalent representation to which one assigns an upper index
\be
{(\p_\a)}^{*}={\bar\p}^\a\ , \qquad \a=1,2\ .
\nonumber
\ee
As usual, spinor indices are raised and lowered with the use of the $SU(2)$--invariant
antisymmetric matrices
\be
\p^\a=\e^{\a\b}\p_\b\ , \quad {\bar\p}_\a=\e_{\a\b} {\bar\p}^\b\ ,
\nonumber
\ee
where $\e_{12}=1$, $\e^{12}=-1$. For spinor bilinears we stick to the notation
\be
\quad \p^2=(\p^\a \p_\a\ ) , \quad
\bar\p^2=(\bar\p_\a \bar\p^\a )\ , \quad \bar\p \p=(\bar\p^\a \p_\a )\ ,
\nonumber
\ee
such that
\be
\p_\a \p_\b=\frac 12 \e_{\a\b} \p^2\ , \qquad \bar\p^\a \bar\p^\b=\frac 12 \e^{\a\b} \bar\p^2\ , \qquad \p_\a \bar\p_\b-\p_\b \bar\p_\a=\e_{\a\b}
(\bar\p \p)\ .
\nonumber
\ee

The Pauli matrices ${{(\s_a)}_\a}^\b$
are taken in the standard form
\be
\s_1=\begin{pmatrix}0 & 1\\
1 & 0
\end{pmatrix}\ , \qquad \s_2=\begin{pmatrix}0 & -i\\
i & 0
\end{pmatrix}\ ,\qquad
\s_3=\begin{pmatrix}1 & 0\\
0 & -1
\end{pmatrix}\ ,
\nonumber
\ee
which obey
\bea
&&
{{(\s_a \s_b)}_\a}^\b +{{(\s_b \s_a)}_\a}^\b=2 \d_{ab} {\d_\a}^\b \ , \quad
{{(\s_a \s_b)}_\a}^\b -{{(\s_b \s_a)}_\a}^\b=2i \e_{abc} {{(\s_c)}_\a}^\b \ ,
\nonumber\\[2pt]
&&
{{(\s_a \s_b)}_\a}^\b=\d_{ab} {\d_\a}^\b +i \e_{abc} {{(\s_c)}_\a}^\b \ , \quad
{{(\s_a)}_\a}^\b {{(\s_a)}_\g}^\r=2 {\d_\a}^\r {\d_\g}^\b-{\d_\a}^\b {\d_\g}^\r\ ,
\nonumber\\[2pt]
&&
{{(\s_a)}_\a}^\b \e_{\b\g} ={{(\s_a)}_\g}^\b \e_{\b\a}\ , \quad \e^{\a\b} {{(\s_a)}_\b}^\g=\e^{\g\b} {{(\s_a)}_\b}^\a \ ,
\nonumber
\eea
where $\e_{abc}$ is the totally antisymmetric Levi-Civit\'a tensor, $\e_{123}=1$. Throughout the text we use the abbriviation
$\bar\p \s_a \p=\bar\p^\a {{(\s_a)}_\a}^\b \p_\b$.

When constructing an $N=4$ supersymmetric extension of the particle model on $AdS_2 \times S^2$ background within the Hamiltonian formalism,
one uses a pair of complex conjugate spinors $(\p_\a, \bar\p^\a)$ in order to parameterize the odd sector of the phase space.
These obey the Poisson bracket
\be
\{ \p_\a, \bar\p^\b \}=-i{\d_\a}^\b \ .
\nonumber
\ee
Conserved charges of an ultimate theory must obey $su(1,1|2)$ superalgebra which is taken in the form
\begin{align}\label{algebra}
&
\{ H,D \}=H\ , && \{ H,K \}=2D\ ,
\nonumber\\[2pt]
&
\{D,K\}=K\ , && \{ \mathcal{J}_a,\mathcal{J}_b \}=\epsilon_{abc} \mathcal{J}_c\ ,
\nonumber\\[2pt]
&
\{ Q_\a, \bar Q^\b \}=-2 i H {\d_\a}^\b\ , &&
\{ Q_\a, \bar S^\b \}=2{{(\s_a)}_\a}^\b \mathcal{J}_a+2iD {\d_\a}^\b-C {\d_\a}^\b\ ,
\nonumber\\[2pt]
&
\{ S_\a, \bar S^\b \}=-2i K {\d_\a}^\b\ , &&
\{ \bar Q^\a, S_\b \}=-2{{(\s_a)}_\b}^\a \mathcal{J}_a+2iD {\d_\b}^\a+C {\d_\b}^\a\ ,
\nonumber\\[2pt]
& \{ D,Q_\a\} = -\frac{1}{2} Q_\a\ , && \{ D,S_\a\} =\frac{1}{2} S_\a\ ,
\nonumber\\[2pt]
&
\{ K,Q_\a \} =S_\a\ , && \{ H,S_\a \}=-Q_\a\ ,
\nonumber\\[2pt]
&
\{ \mathcal{J}_a,Q_\a\} =\frac{i}{2} {{(\s_a)}_\a}^\b Q_\b\ , && \{ \mathcal{J}_a,S_\a\} =\frac{i}{2} {{(\s_a)}_\a}^\b S_\b\ ,
\nonumber\\[2pt]
& \{ D,\bar Q^\a \} =-\frac{1}{2} \bar Q^\a\ , && \{ D,\bar S^\a\} =\frac{1}{2} \bar S^\a\ ,
\nonumber\\[2pt]
& \{K,\bar Q^\a\} =\bar S^\a\ , && \{ H,\bar S^\a\} =-\bar Q^\a\ ,
\nonumber\\[2pt]
&
\{\mathcal{J}_a,\bar Q^\a\} =-\frac{i}{2} \bar Q^\b {{(\s_a)}_\b}^\a\ , && \{ \mathcal{J}_a,\bar S^\a\} =-\frac{i}{2}
\bar S^\b {{(\s_a)}_\b}^\a\ .
\nonumber
\end{align}
Here $C$ is the central charge. The missing Poisson brackets prove to vanish.

\end{document}